\documentclass[twocolumn,showpacs,preprintnumbers,amsmath,amssymb,prl]{revtex4}

\usepackage{graphicx}
\usepackage{dcolumn}
\usepackage{bm}
\usepackage{amssymb}

\begin{document}


\title{Magnetic Moment Coupling to Circularly Polarized Photons}

\author{O.V. Kibis}\email{Oleg.Kibis@nstu.ru}
\thanks{{\tt www.kibis.ru}}

\affiliation{Department of Applied and Theoretical Physics,
Novosibirsk State Technical University, Karl Marx Avenue 20,
630092 Novosibirsk, Russia}


\begin{abstract}
Exact stationary solutions of the wave equation are obtained to
describe the interaction between magnetic moment of elementary
particle and circularly polarized photons. The obtained solutions
substantially modify the conventional model of field-matter
interaction. It follows from them that magnetic moment couples to
photons, and this coupling leads to bound particle-photon states
with different energies for different orientations of magnetic
moment. As a consequence, the interaction splits particle states
differing by directions of total angular momentum. Stationary spin
splitting, induced by photons, and concomitant effects can be
observed for particles exposed to a laser-generated circularly
polarized electromagnetic wave.
\end{abstract}


\maketitle

The interaction between magnetic moment of elementary particle and
electromagnetic field is one of fundamental interactions in the
nature. The quantum theory to describe one was elaborated at the
dawn of quantum mechanics and has taken place in textbooks (e.g.,
\cite{Landau_3,Landau_4}). However, this theory deals with
electromagnetic field in the framework of classical
electrodynamics. A theory, capable of describing the interaction
for quantized electromagnetic field, was unknown up to the
present. This gap in the fundamental area of quantum physics
arises from formidable mathematical obstacles to find solutions of
the wave equation in the case of intensive quantized field. A
small parameter, that would help to find the solution as a series
expansion, is absent in the problem. As a consequence, the
standard quantum-electrodynamic methodology, based on the
perturbation theory, is not applicable here. Therefore there is no
other way to work out the problem but solve this equation
accurately. Unfortunately, a regular method to solve the wave
equation is unknown, and success in seeking exact solutions for
special cases depends on the Fortune. That is the reason why the
problem was not elaborated before now. The author of the Letter
was lucky to find exact stationary solutions of the wave equation
describing the interaction between magnetic moment of particle and
intensive quantized circularly polarized electromagnetic field.
Since the obtained solutions substantially modify the conventional
model of particle-field interaction, they are of broad interest to
the physics community.

First of all, let us consider the interaction between a
monochromatic circularly polarized electromagnetic wave and an
electrically neutral particle with a magnetic moment $\mu$ and the
spin-1/2 (for example, neutron). The interaction Hamiltonian,
written in the conventional form, is given by $\hat{\cal
H}_{{\mathrm{int}}}=-\hat{\bm{\mu}}\mathbf{H}$, where
$\hat{\bm{\mu}}=\mu\hat{\bm{\sigma}}$ is operator of the magnetic
moment, $\hat{\bm{\sigma}}$ is the Pauli matrix operator, and
$\mathbf{H}$ is classical magnetic field of the wave
\cite{Landau_3}. Considering the problem within the standard
quantum-field approach \cite{Landau_4}, the field, $\mathbf{H}$,
should be replaced with the field operator, $\hat{\mathbf{H}}$.
Assuming the wave to be clockwise-polarized, this operator can be
written as
\begin{equation}\label{H}
\hat{\mathbf{H}}=i\widetilde{H}_0\left[\mathbf{e}_z\times\left(\mathbf{e}_+\hat{a}e^{ik_0
z}-\mathbf{e}_-\hat{a}^\dagger e^{-ik_0 z}\right)\right]\,,
\end{equation}
where $\widetilde{H}_0=\sqrt{2\pi\hbar\omega_0/V}$, $\omega_0$ is
frequency of the wave, $V$ is volume,
$\mathbf{e}_{\pm}=(\mathbf{e}_{x}\pm i\mathbf{e}_{y})/\sqrt{2}$
are polarization vectors, $\mathbf{e}_{x,y,z}$ are unit vectors
directed along the $x,y,z$-axes, $k_0=\omega_0/c$ is wave vector
assumed to be directed along the $z$-axis, $\hat{a}$ and
$\hat{a}^\dagger$ are operators of destruction and creation of
photons in the wave, respectively \cite{Landau_4}. The replacement
leads to the Hamiltonian of the particle-photon system
\begin{equation}\label{PFH}
\hat{{\cal
H}}=\frac{\hat{\mathbf{p}}^2}{2m}+\hbar\omega_0\hat{a}^\dagger\hat{a}
-\sqrt{2}\mu\widetilde{H}_0\left(\hat{\sigma}_+\hat{a}e^{ik_0
z}+\hat{\sigma}_-\hat{a}^\dagger e^{-ik_0z}\right),
\end{equation}
where the first term on the right-hand side describes kinetic
energy of the particle, the second term corresponds to field
energy, the third term is the interaction Hamiltonian, $\hat{\cal
H}_{{\mathrm{int}}}$, rewritten in quantum-field form,
$\hat{\mathbf{p}}$ is momentum operator of the particle, $m$ is
mass of the particle, and $\hat{\sigma}_{\pm}=(\hat{\sigma}_x\pm
i\hat{\sigma}_y)/2$ are step-up and step-down operators for the
$z$-projection of the particle spin, $\mathbf{S}$.

To describe the particle-photon system, let us use the notation
$|S_z,N\rangle$ which indicates that the particle is in quantum
state with the spin projection $S_z=+1/2$ or $S_z=-1/2$ and the
wave is in quantum state with the photon occupation number $N$.
Then the exact stationary solutions of the wave equation with the
Hamiltonian (\ref{PFH}), $\psi_{+1/2,N_0}$ and $\psi_{-1/2,N_0}$,
can be written as
\begin{eqnarray}\label{Psi}
|\psi_{\pm1/2,N_0}\rangle=\frac{e^{i\mathbf{k}\mathbf{r}}}{\sqrt{V}}
\left[\sqrt{\frac{\Omega_\pm+\omega_\pm}{2\,\Omega_\pm}}\,|\pm1/2,N_0\rangle\right.\,\,\,\,\,\,\,\,\,
\,\,\,\,\,\,\,\,\,\,\,\,\,\nonumber\\
\pm\left.e^{\mp ik_0
z}\sqrt{\frac{\Omega_\pm-\omega_\pm}{2\,\Omega_\pm}}\,|\mp1/2,N_0\pm1\rangle\right]e^{-i\varepsilon_{\pm1/2,N_0}t/\hbar},
\,\,\,\,\,\,\,\,
\end{eqnarray}
where $N_0$ is photon occupation number of the unperturbed wave,
$\mathbf{k}$ is wave vector of the particle, $\mathbf{r}$ is
radius-vector of the particle, energies of the particle-photon
system, $\varepsilon_{+1/2,N_0}$ and $\varepsilon_{-1/2,N_0}$, are
given by
\begin{equation}\label{Energy}
\varepsilon_{\pm1/2,N_0}=\frac{\hbar^2k^2}{2m}
+N_0\hbar\omega_0\pm\frac{\hbar\omega_\pm}{2}\mp\frac{\hbar\Omega_\pm}{2}\,,
\end{equation}
frequencies $\Omega_+$ and $\Omega_-$ are
$$\Omega_\pm=\sqrt{8(N_0+1/2\pm1/2)(\mu\widetilde{H}_0/\hbar)^2+\omega_\pm^2}\,,$$
and $\omega_\pm=\omega_0(1-\hbar k_z/mc\pm\hbar k_0/2mc)\,$. The
subscript indexes in Eqs.~(\ref{Psi})--(\ref{Energy}) indicate
genesis of the bound particle-photon states, i.e. the state
$|\psi_{\pm1/2,N_0}\rangle$ turns into the state
$|\pm1/2,N_0\rangle$ when the particle-photon interaction vanishes
($\mu=0$). The solutions (\ref{Psi})--(\ref{Energy}) can be easily
verified by direct substitution into the wave equation
$i\hbar\partial\psi_{\pm1/2,N_0}/\partial t=\hat{{\cal
H}}\psi_{\pm1/2,N_0}$ with the Hamiltonian (\ref{PFH}), keeping in
mind the trivial relations \cite{Landau_3,Landau_4}
\begin{eqnarray}
\hat{\sigma}_{\pm}\,|\mp1/2,N\rangle&=&|\pm1/2,N\rangle\,,\,\,\,\,
\hat{\sigma}_{\pm}\,|\pm1/2,N\rangle\,=0\,\,,\nonumber\\
\hat{a}^{\dagger}\,|\pm1/2,N\rangle&=&\sqrt{N+1}\,|\pm1/2,N+1\rangle\,,\nonumber\\
\hat{a}\,|\pm1/2,N\rangle&=&\sqrt{N}\,|\pm1/2,N-1\rangle\,.\nonumber
\end{eqnarray}
The first term on the right-hand side of Eq.~(\ref{Energy}) is
energy of noninteracting particle, the second term is energy of
noninteracting photons, and other terms arise from the
particle-photon interaction. This implies that the ground energy,
$\varepsilon_{+1/2,N_0}$, is less than an energy of the
noninteracting particle-photon system. Therefore the considered
interaction results in stable particle coupling to photons. It
follows from the equality
\begin{equation}\label{Sigma}
\langle\psi_{\pm1/2,N_0}|\hat{\bm{\sigma}}|\psi_{\pm1/2,N_0}\rangle=
\pm\omega_\pm\mathbf{e}_{z}/\Omega_\pm\,
\end{equation}
that the states $\psi_{+1/2,N_0}$ and $\psi_{-1/2,N_0}$ correspond
to mutually opposite orientations of averaged particle spin. Thus
the difference in energy of these states,
$\Delta\varepsilon=\varepsilon_{-1/2,N_0}-\varepsilon_{+1/2,N_0}$,
should be interpreted as spin splitting induced by photons. Let us
emphasize that the Hamiltonian (\ref{PFH}) holds true only for the
wave intensive enough. If the photon occupation number of the
unperturbed wave, $N_0$, is small, intensities of the particle
interaction with photons from the wave and with vacuum states of
other photons are comparable. Therefore for small occupation
numbers of clockwise-polarized photons in the wave, the
Hamiltonian (\ref{PFH}) should be supplemented with terms
describing the interaction between magnetic moment and vacuum
states of counterclockwise-polarized photons with the same wave
vector $k_0$. With these terms accounted, the photon-induced spin
spitting $\Delta\varepsilon$ vanishes for $N_0=0$, as expected. In
what follows we shall be to assume the wave to be intensive
($N_0\gg1$) and the particle to be nonrelativistic
($\omega_\pm\approx\omega_0$). Then the frequencies $\Omega_\pm$
in Eqs.~(\ref{Psi})--(\ref{Energy}) can be replaced with the
frequency
\begin{equation}\label{Omega}
\Omega=\sqrt{(2\mu H_0/\hbar)^2+\omega_0^2}\,,
\end{equation}
where $H_0=\sqrt{2N_0}\widetilde{H}_0$ is classical amplitude of
magnetic field of the wave. Thus the circularly polarized
electromagnetic wave leads to the stationary spin splitting
\begin{equation}\label{Delta}
\Delta\varepsilon=\sqrt{(2\mu
H_0)^2+(\hbar\omega_0)^2}-\hbar\omega_0\,,
\end{equation}
which cannot be described by the conventional model of
particle-field interaction \cite{Landau_3,Landau_4}, based on the
classical electrodynamics. It follows from the aforesaid that this
splitting should be considered as novel quantum-field effect
arising from the dressing of particle by circularly polarized
photons. For neutrons exposed to the wave generated by a modern
petawatt laser, the splitting (\ref{Delta}) may approach the
electron-Volt level.

The problem solved above for the spin-1/2 can be generalized for a
particle with arbitrary total angular momentum, $\mathbf{J}$. In
this case, the magnetic moment operator can be written as
$\hat{\bm{\mu}}=(\mu/J)\hat{\mathbf{J}}$, where $\hat{\mathbf{J}}$
is operator of total angular momentum \cite{Landau_3}.
Substitution of this magnetic moment operator into the interaction
Hamiltonian, $\hat{\cal H}_{{\mathrm{int}}}$, results in replacing
the operators $\hat{\sigma}_\pm$ with the operators
$\hat{J}_\pm=(\hat{J}_x\pm i\hat{J}_y)/2J$ in the complete
Hamiltonian of the particle-photon system (\ref{PFH}). The regular
procedure to solve accurately the wave equation with the modified
Hamiltonian (\ref{PFH}) is as follows. The solutions,
$\psi_{j,N_0}$, should be sought in the form
\begin{eqnarray}\label{PsinP}
|\psi_{j,N_0}\rangle&=&
\frac{e^{i\mathbf{k}\mathbf{r}}}{\sqrt{V}}\sum_{n=-J}^{J}C_{j,N_0}^{(n)}e^{i(n-j)k_0
z}|n,N_0+j-n\rangle\nonumber\\
&\times&e^{-i\varepsilon_{j,N_0}t/\hbar}\,,
\end{eqnarray}
where the notation $|J_z,N\rangle$ indicates that the particle is
in quantum state with the total angular momentum projection $J_z$
and the wave is in quantum state with the photon occupation number
$N$. Keeping in mind the relation \cite{Landau_3}
\begin{eqnarray}
\langle J_z,N|\,\hat{J}_{+}\,|J_z-1,N\rangle&=&\langle
J_z-1,N|\,\hat{J}_{-}\,|J_z,N\rangle\nonumber\\
&=&\sqrt{(J+J_z)(J-J_z+1)}/2J\,,\nonumber
\end{eqnarray}
substitution of the function (\ref{PsinP}) into the wave equation
with the modified Hamiltonian (\ref{PFH}) results in the system of
$2J+1$ homogeneous algebraic equations
\begin{eqnarray}
\Bigg[\frac{\hbar^2k^2}{2m}
+N_0\hbar\omega_0+(j-n)\hbar\omega_{j-n}-\varepsilon_{j,N_0}\Bigg]C_{j,N_0}^{(n)}\nonumber\\
-\,\frac{\mu\widetilde{H}_0}{\sqrt{2}J}\left[\sqrt{(N_0+j-n)(J+n+1)(J-n)}\,C_{j,N_0}^{(n+1)}\right.
\nonumber\\
+\left.\sqrt{(N_0+j-n+1)(J+n)(J-n+1)}\,C_{j,N_0}^{(n-1)}\right]=0\,,\nonumber\\
n=-J,-J+1,...,J-1,J\,,\nonumber
\end{eqnarray}
where $\omega_l=\omega_0(1-\hbar k_z/mc+l\hbar k_0/2mc)\,$. The
well known procedure of solving such an algebraic system leads to
$2J+1$ sets of solutions,
$\left\{\varepsilon_{j,N_0},\,C_{j,N_0}^{(n)}\right\}$, which
define $2J+1$ wave functions (\ref{PsinP}). The parameter $j$,
undefined before, should be specified independently for each of
the sets in order to turn the bound particle-photon state
$|\psi_{j,N_0}\rangle$ into the state $|j,N_0\rangle$ when the
particle-photon interaction vanishes (i.e. for $\mu=0$). It
appears that this parameter is equal to different values
$-J,-J+1,...,J-1,J$ for different wave functions (\ref{PsinP}) and
should be interpreted as the $z$-projection of total angular
momentum of the noninteracting particle. As expected, for
particles with the total angular momentum $J=1/2$ the described
procedure leads to the solutions (\ref{Psi})--(\ref{Energy}).
Omitting relativistically small terms ($\omega_l\approx\omega_0)$
and keeping in mind that the wave is intensive $(N_0\gg2J)$, the
energy of the particle-photon system for arbitrary $J$ can be
written as
\begin{eqnarray}\label{EnergyP}
\varepsilon_{j,N_0}&=&\frac{\hbar^2k^2}{2m}+(N_0+j)\hbar\omega_0
-j\sqrt{(\mu H_0/J)^2+(\hbar\omega_0)^2}\,,\nonumber\\
j&=&-J,-J+1,...,J-1,J\,.
\end{eqnarray}
Thus the interaction leads to different energies for different
$z$-projections, $j$, of total angular momentum, $\mathbf{J}$. It
should be noted that Eq.~(\ref{EnergyP}) is applicable also to
atoms which can formally be considered as electrically neutral
particles, if a wave field is much weaker than an intra-atomic
field. In this case the wave does not change substantially an
intra-atomic structure, and $\mu$ can be interpreted as known
effective magnetic moment of unperturbed atom. Then the total
angular momentum, $\mathbf{J}$, includes a nuclear spin, electron
spins and orbital angular momentums of electrons. As a result, the
atom-photon interaction splits atom energy levels totally. If the
wave is generated by an usual laser with the wavelength $\sim\mu$m
and the intensity of radiation $\sim10^{8}$~W/cm$^2$, the
Eq.~(\ref{EnergyP}) can be used to find the photon-induced spin
splitting of the ground $1S$-state in a hydrogen atom. Neglecting
the small nuclear magnetic moment, the splitting is described by
Eq.~(\ref{Delta}), where $\mu$ is equal approximately to the
electron Bohr magneton. The calculation leads to the splitting
value $\Delta\varepsilon\sim10^{-4}$~eV that is tens times as
large as the usual hyperfine splitting of the $1S$-state
\cite{Landau_3}. Naturally, besides the energy splitting in
isolated atoms, circularly polarized photons can also induce the
energy gap opening in molecules and different condensed-matter
structures, that will be analyzed elsewhere.

To complete the analysis, let us consider the photon coupling to
an electrically charged particle with the spin-1/2 (for example,
electron). It is well known that interaction between a circularly
polarized electromagnetic wave and a free charged particle leads
to rotation of the particle \cite{Landau_2}. This rotation causes
the additional (spin-orbit) interaction between magnetic moment of
the particle and electric field of the wave. As a consequence, we
need to modify the Hamiltonian (\ref{PFH}) for this case. Firstly,
the Hamiltonian should be supplemented with a term describing the
spin-orbit interaction. In the framework of classical
electrodynamics the term has the form \cite{Landau_4}
\begin{equation}\label{HSO}
\hat{{\cal
H}}_{so}=-(\mu_a+\mu_B/2)\hat{\bm{\sigma}}[{\mathbf{E}}\times\hat{\mathbf{v}}/c],
\end{equation}
where $\mu_a$ is anomalous magnetic moment of the particle,
$\mu_B=e\hbar/2mc$ is Bohr magneton of the particle, $e$ is
electric charge of the particle,
$\hat{\mathbf{v}}=\hat{\mathbf{p}}/m$ is operator of particle
velocity, and ${\mathbf{E}}$ is electric field of the wave.
Secondly, the momentum operator $\hat{\mathbf{p}}$ should be
replaced with the operator $\hat{\mathbf{p}}-e{\mathbf{A}}/c$,
where ${\mathbf{A}}$ is vector potential of the wave. Thirdly, the
classical fields, ${\mathbf{A}}$ and ${\mathbf{E}}$, should be
replaced with the field operators, $\hat{\mathbf{A}}$ and
$\hat{\mathbf{E}}$, respectively. Using the well known expressions
for these field operators \cite{Landau_4}, the modified
Hamiltonian (\ref{PFH}) can be written as $\hat{{\cal
H}}=\hat{{\cal H}}^{\prime}+\hat{{\cal H}}^{\prime\prime}$, where
\begin{eqnarray}
\hat{{\cal H}}^{\prime}&=&\hat{\mathbf{p}}^2/2m
-\sqrt{2}\mu\widetilde{H}_0\left(\hat{\sigma}_+\hat{a}e^{ik_0
z}+\hat{\sigma}_-\hat{a}^\dagger
e^{-ik_0z}\right)+\hat{a}^\dagger\hat{a}\nonumber\\
&\times&\left[\hbar\omega_0+\left(e\widetilde{H}_0^2/mc\omega_0^2\right)\Big(ec+(2\mu_a+\mu_B)\omega_0
\hat{\sigma}_z\Big)\right]\,,\nonumber\\
\hat{{\cal
H}}^{\prime\prime}&=&-\left(\widetilde{H}_0/m\right)\left(e/\omega_0+\mu_a/c+
\mu_B/2c\right)\nonumber\\
&\times&\left(\hat{p}_+\hat{a}e^{ik_0 z}+\hat{p}_-\hat{a}^\dagger
e^{-ik_0z}\right)\,,\nonumber
\end{eqnarray}
and $\hat{p}_{\pm}=(\hat{p}_x\pm i\hat{p}_y)/\sqrt{2}$. Exact
solutions of the wave equation with the Hamiltonian $\hat{{\cal
H}}^{\prime}$ can be found in the form (\ref{PsinP}) with $J=1/2$
by invoking the procedure described above. As to the Hamiltonian
$\hat{{\cal H}}^{\prime\prime}$, it can be accounted by using the
standard perturbation theory \cite{Landau_3} which is applicable
for states (\ref{PsinP}) with small wave vector components $k_x$
and $k_y$. In this way we shall obtain expressions for energy
levels $\varepsilon_{+1/2,N_0}$ and $\varepsilon_{-1/2,N_0}$. As a
result, the photon-induced spin splitting,
$\Delta\varepsilon=\varepsilon_{-1/2,N_0}-\varepsilon_{+1/2,N_0}$,
is given by
\begin{equation}\label{DeltaCharge}
\Delta\varepsilon=2\sqrt{[\mu H_0]^2-\left[\frac{\hbar
eH_0^2}{mc}\right]\left[\mu_a+\frac{\mu_B}{2}\right]
+\left[\frac{\hbar\omega_0}{2}\right]^2}-\hbar\omega_0
\end{equation}
for $p_0/mc\ll1$, where $p_0=eH_0/\omega_0$ is momentum of the
rotating particle \cite{Landau_2}. Since magnetic moment of the
particle is $\mu=\mu_a+\mu_B$, the splitting (\ref{DeltaCharge})
for charged particles in vacuum has the form
\begin{equation}\label{DeltaV}
\Delta\varepsilon=\sqrt{(2\mu_a
H_0)^2+(\hbar\omega_0)^2}-\hbar\omega_0
\end{equation}
and depends only on anomalous part of magnetic moment, $\mu_a$.
Considering free electrons in condensed matter, their mass $m$ in
Eq.~(\ref{DeltaCharge}) should be interpreted as effective
electron mass, $m_0^\ast$, while the Bohr magneton
$\mu_B=e\hbar/2m_0c$ depends on electron mass in vacuum, $m_0$. In
this case the sign of $\Delta\varepsilon$ depends on $m_0^\ast$.
Neglecting the small quantity $\mu_a$ for electrons, we obtain
from Eq.~(\ref{DeltaCharge}) that for $m_0<m_0^\ast$ the ground
electron state in condensed matter is $\varepsilon_{+1/2,N_0}$,
and for $m_0>m_0^\ast$ one is $\varepsilon_{-1/2,N_0}$.

If the particle rotation induced by the wave is suppressed, the
interaction (\ref{HSO}) does not influence on the spin splitting.
Then the interaction Hamiltonian can be written in the same form,
$\hat{\cal H}_{{\mathrm{int}}}=-\hat{\bm{\mu}}\hat{\mathbf{H}}$,
as for uncharged particles. As a result, in this case the
expressions (\ref{Delta}) and (\ref{EnergyP}), obtained before for
electrically neutral particles, can be used for charged particles
as well. Such a suppression takes place for confined charged
particles, including electrons in atoms and nanostructures, as
well as for free electrons in condensed matter for
$\omega_0\tau\ll1$, where $\tau$ is electron mean free time. For
electrons the splitting (\ref{DeltaV}) is much less than the
splitting (\ref{Delta}) because of the ratio
$\mu_a/\mu\sim10^{-3}$. Therefore electron systems, where the
rotation is suppressed, are most suitable for observation of the
discussed effect.

Formally, the results described above are obtained for
nonrelativistic particles. To describe the relativistic case, we
have to start from the Hamiltonian, $\hat{{\cal
H}}=\hbar\omega_0\hat{a}^\dagger\hat{a}+\hat{{\cal
H}}_D(\hat{\mathbf{p}}-e\hat{\mathbf{A}}/c)$, based on the Dirac
Hamiltonian $\hat{{\cal H}}_D(\hat{\mathbf{p}})$ \cite{Landau_4},
that will be done elsewhere. Running ahead, it should be noted
that the relativistic analysis predicts novel effects (for
instance, the increasing of particle mass, arising from the photon
dress of particle), but leads to the same above-described
expressions in the limiting case of small particle velocities, as
expected.

Finalizing the Letter, let us duscuss possible observable
consequences of the particle-photon coupling described by
Eqs.~(\ref{Psi})--(\ref{Delta}). The first effect of the
photon-induced spin splitting (\ref{Delta}) is magnetization of
particles exposed to the wave. It follows from Eq.~(\ref{Sigma})
that the particle, being in the ground state
$\varepsilon_{+1/2,N_0}$, is spin-polarized along the $z$-axis.
This implies that an equilibrium gas of particles, exposed to the
circularly polarized wave, will be spin-polarized along angular
momentum vector of photons. In the particular case of
nondegenerate gas, the magnetization vector can be written as
${\mathbf{M}}=\left(\mu n\omega_0/\Omega\right)\tanh
\left(\Delta\varepsilon/2T\right)\mathbf{e}_{z}$, where $T$ is
temperature, and $n$ is density of the gas.

The second effect is optical transitions with frequencies
different from the wave frequency, $\omega_0$. That effect arises
from interaction between the bound particle-photon states
(\ref{Psi}) and free photons of other kinds. Nonzero magnetodipole
moments for the transitions are given by
\begin{eqnarray}\label{MM}
|\langle\psi_{\pm1/2,N_0\mp1}|\mathbf{e}_{\mp}\hat{\bm{\mu}}|\psi_{\pm1/2,N_0}\rangle|&=&
|\mu|\frac{\sqrt{\Omega^2-\omega^2_0}}{\sqrt{2}\Omega}\,,\nonumber\\
|\langle\psi_{\mp1/2,N_0}|\mathbf{e}_{\mp}\hat{\bm{\mu}}|\psi_{\pm1/2,N_0}\rangle|&=&
|\mu|\frac{\Omega+\omega_0}{\sqrt{2}\Omega}\,, \nonumber\\
|\langle\psi_{\mp1/2,N_0\pm2}|\mathbf{e}_{\pm}\hat{\bm{\mu}}|\psi_{\pm1/2,N_0}\rangle|&=&
|\mu|\frac{\Omega-\omega_0}{\sqrt{2}\Omega}\,, \nonumber\\
|\langle\psi_{\mp1/2,N_0\pm1}|\mathbf{e}_{z}\hat{\bm{\mu}}|\psi_{\pm1/2,N_0}\rangle|&=
&|\mu|\frac{\sqrt{\Omega^2-\omega^2_0}}{2\Omega}\,.
\end{eqnarray}
As a result, transitions between the states (\ref{Psi}) with
nonzero matrix elements (\ref{MM}) can be accompanied by
magnetodipole emission and absorption of electromagnetic
radiation. It follows from the expressions (\ref{MM}) and
(\ref{Energy}) that there are the three new transition
frequencies, $\Omega,\,\Omega-\omega_0,\,\Omega+\omega_0\,$.
Allowed optical transitions with these frequencies from the ground
state $\varepsilon_{+1/2,N_0}$ are pictured in Fig.~1. by arrows.
\begin{figure}[th]
\includegraphics[width=0.48\textwidth]{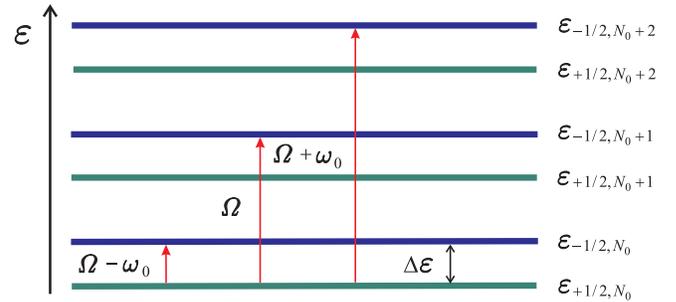}
\caption{Schematics of optical transitions between bound
particle-photon states.}\label{fig1}
\end{figure}
It should be stressed that the frequency (\ref{Omega}) depends on
the magnetic field amplitude, $H_0$, that leads to dependence of
the transition frequencies on a wave intensity. Besides
transitions with the above-mentioned new frequencies, there are
optical transitions with the wave frequency, $\omega_0$. They are
described by the first of the matrix elements (\ref{MM}). These
transitions change the photon occupation number of the wave by one
and should be interpreted as scattering of photons by the
particle. It should be reminded that optical transitions are
accompanied by momentum transfer to particles from photons, that
can change mechanical energy of particles (the quantum recoil
effect). As a consequence, the above-mentioned frequencies
describe the optical transitions accurately when the quantum
recoil can be neglected. This takes place, particularly, for
confined particles. Let us stress that the bound particle-photon
states, $\psi_{\pm1/2,N}$, can be classified by the $z$-projection
of angular momentum of the particle-photon system, $l_z=N\pm1/2$.
As expected, the nonzero matrix elements (\ref{MM}) correspond to
transitions between states (\ref{Psi}) with projections $l_z$
different by $-1,0,1$.

It should be reminded that the electromagnetic wave has been
assumed to have the clockwise polarization. If the wave is
counterclockwise-polarized, the operators $\hat{\sigma}_+$ and
$\hat{\sigma}_-$ ($\hat{J}_+$ and $\hat{J}_-$) in the Hamiltonian
(\ref{PFH}) should be permuted. In this case the expressions
following from the Hamiltonian retain their form but particle spin
(or, generally, total angular momentum of particle) changes its
direction to the opposite. As to other kinds of photon
polarization, it can be shown that magnetic moment couples to
elliptically polarized photons weaker than to circularly polarized
ones, and in the limiting case of linear polarization the coupling
vanishes.

It follows from the aforesaid that the predicted particle coupling
to photons is fundamental quantum effect unexplored before.
Certainly, the presented first analysis does not exhaust the
problem entirely but forms a basis for further experimental and
theoretical studies.

\end{document}